\documentclass{elsart}

\usepackage{amssymb}
\usepackage{graphicx}

\begin{document}

\begin{frontmatter}

\title{Collaborative filtering based on multi-channel diffusion}

\author[1]{Ming-Sheng Shang}
\author[2]{Ci-Hang Jin}
\author[2]{Tao Zhou}
\author[1,2]{Yi-Cheng Zhang}

\address[1]{Lab of Information Economy and Internet Research,University of Electronic Science and Technology, 610054 Chengdu, China}
\address[2]{Department of Physics, University of Fribourg, Chemin du
Mus\'ee 3, CH-1700 Fribourg, Switzerland}

\begin{abstract}
In this paper, by applying a diffusion process, we propose a new
index to quantify the similarity between two users in a user-object
bipartite graph. To deal with the discrete ratings on objects, we
use a multi-channel representation where each object is mapped to
several channels with the number of channels being equal to the
number of different ratings. Each channel represents a certain
rating and a user having voted an object will be connected to the
channel corresponding to the rating. Diffusion process taking place
on such a user-channel bipartite graph gives a new similarity
measure of user pairs, which is further demonstrated to be more
accurate than the classical Pearson correlation coefficient under
the standard collaborative filtering framework.
\end{abstract}


\end{frontmatter}

\section{Introduction}

With the rapid growth of the Internet \cite{NJP08} and the
World-Wide-Web \cite{Broder00}, a huge amount of data and resource
is created and available for the public. This, however, may result
in a dilemma problem. On the one hand, the unprecedented growth of
available information has brought us into the world of many
possibilities: people may choose from thousands of movies, millions
of books, and billions of web pages; on the other hand, the amount
of information is increasing more quickly than our personal
processing abilities and therefore evaluations of all alternatives
are not feasible at all. In consequence, it is vital to
automatically extract the hidden information and make personalized
recommendations.

A lot of work has been done in this field. A landmark is the use of
\emph{search engine} \cite{Brin1998,Kleinberg1999}. However, a
search engine could only find the relevant web pages according to
the input keywords and returns the same results regardless of users'
habits and tastes. Another landmark is the so-called
\emph{recommender system} \cite{Resnick1997}, which is essentially
an information filtering technique that attempts to find out objects
likely to be interesting to the target users. Due to its
significance for economy and society, the design of efficient
recommendation algorithms has become a common focus for computer
science, mathematics, marketing practices, management science and
physics (see the review articles
\cite{Adomavicius05,Herlocker04,LiuRev} and the references therein).

Various kinds of recommendation algorithms have been proposed,
including the content-based analysis \cite{Pazzani2007}, the
spectral methods \cite{Maslov00,Goldberg2001}, the heat conduction
algorithm \cite{Zhang2007a}, the opinion diffusion algorithm
\cite{Zhang2007b}, the network-based inference
\cite{Zhou2007,Zhou2008}, the latent semantic model
\cite{Hofmann2004}, the latent Dirichlet allocation \cite{Blei2003},
the iterative self-consistent refinement \cite{Ren2008}, and so on.
Among them, collaborative filtering (CF) is one of the earliest and
the most successful algorithms underlies recommender systems
\cite{Schafer2007}. A latent assumption of CF approach is that, in a
social network, those who agreed in the past tend to agree again in
the future. The most commonly used algorithmic framework of CF
consists of two steps: firstly to identify the neighborhood of each
user by computing similarities between all pairs of users based on
their historical preferences, and then to predict by integrating
ratings of target user's neighbors.

Algorithms within this framework differ in the definition of
similarity, the formulation of neighborhoods and the computation of
predictions. The most crucial ingredient in determining the accuracy
of CF is how to properly quantify the similarity between user pairs
\cite{Fouss2007}. In the simplest case, a recommender system can be
well described by a bipartite user-object network \cite{Zhou2007},
where the relations between users and objects are binary: either
presence or absence. For example, in \emph{Amazon.com} users are
connected with books they purchased \cite{Linden2003}, and in
\emph{audioscrobbler.com} listeners are connected with the music
groups they collected \cite{Lambiotte}. Under this bipartite case,
the cosine similarity \cite{Salton1983} is the most widely used
index to quantify the proximity of user tastes. Recently, some new
similarity indices are proposed and shown to be more accurate than
the cosine similarity, including the random-walk-based similarities
\cite{Fouss2007,Gobel1974}, the diffusion-based similarities
\cite{Huang2004,Liu2009,LiuJG2009}, the transferring similarity
\cite{Sun2009}, and so on. In addition, Fouss \emph{et al.}
\cite{Fouss2007} have demonstrated that some classical similarity
indices, such as the Katz index \cite{Katz1953} and the matrix
forest similarity \cite{Chebotarev1997}, can give really good
recommendations under the framework of CF. However, most of those
indices are not easily to be exploited in measuring the user
similarity of rating systems, where, instead of the simply binary
correlations, users can vote objects by different ratings. For
example, In \emph{Yahoo music}, \emph{Netflix.com} and
\emph{MovieLens}, people votes songs by discrete ratings from 1 to
5. In such rating systems, the \emph{Pearson correlation
coefficient} is the most widely used similarity measure
\cite{Adomavicius05}. In the calculation of the Pearson coefficient,
each rating is treated as a number. Taking again the Yahoo Music as
an example, the five ratings, from 1 to 5, corresponding to ``Never
play again", ``It is ok", ``I like it", ``I love it", and ``Can't
get enough", and it is clear that the distance of feelings between
``Never play again" and ``It is ok" is much larger than the distance
between ``I love it" and ``Can't get enough", however, when the
ratings are treated as numbers, rich information gets lost and the
distance between two neighboring ratings is supposed to be the same
(in this example, it is one).

\begin{figure}
  \begin{center}
       \center \includegraphics[width=4.0in]{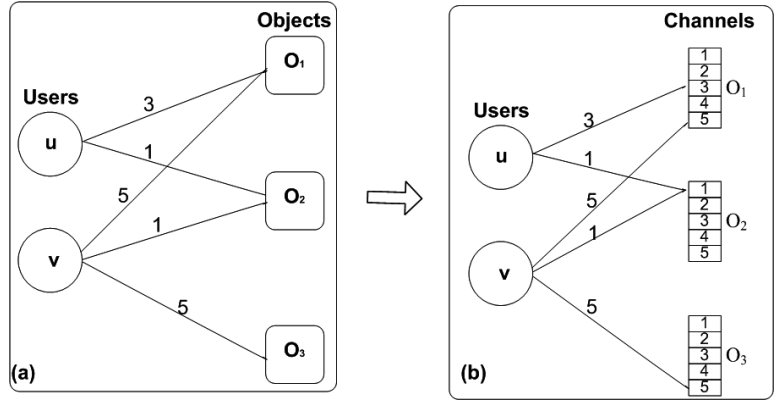}
       \caption{Illustration of the two representations of a five-rating system. Plot (a)
shows a routine representation where weights on edges denote the
corresponding ratings. Plot (b) describes the multi-channel model
where every object is divided into five channels, each of which
represents a rating. User who votes an object is connected to the
channel corresponding to the rating.}\label{fig1}
 \end{center}
\end{figure}

To best keep the original information, we divide every object into
several channels, each of which represents a certain rating. Since
most of the currently used recommendation engines adopt a
five-rating system, this division will not bring much extra
computational complexity. Figure 1 illustrates such a division for a
five-rating system: Figure 1(a) is the routine representation with
each object denoted by a node and the ratings are assigned to the
corresponding edges, and Figure 1(b) is the new representation where
each object is denoted by five channels corresponding to the five
ratings. Under this representation, one can apply the diffusion
process, usually only used in the bipartite version in the past
\cite{Zhang2007b}, to the multi-channel systems. In this paper, to
get the user similarity, we use a network-based resource-allocation
method, which can be considered as a two-step diffusion process and
thus much faster than the one based on a certain convergent
condition \cite{Zhang2007b}. We then use this user similarity to
predict ratings under the standard framework of collaborative
filtering. We test this algorithm on two benchmark data sets,
\emph{MovieLens} and \emph{Netflix}, the results demonstrate its
advantage compared with the standard collaborative filtering
adopting Pearson coefficient. This study indicates a strong
potential of applying physical process to target one of the central
scientific problems in the modern information science---how to
automatically extract hidden information.

\section{Method}

\begin{table}
\caption{Comparison of the two similarity indices on
\emph{MovieLens} and \emph{Netflix}. The probe contains $10\%$ of
the total data, namely $p=10$. All the number are obtained by
averaging over five runs, each of which has an independently random
division of training set and probe.} \label{tab2}
\begin{center}
\begin{tabular}{cccc}
\hline\hline
dataset & similarity index & RMSE & MAE \\
\hline
{ } & diffusion-based & $0.9479$ & $0.7415$ \\
\emph{MovieLens} & Pearson   & $1.0259$ & $0.7805$\\
\hline
{ } & diffusion-based & $0.9406 $ & $0.7303$\\
\emph{NetFlix} & Pearson   & $1.0441$ & $0.7858$\\
\hline\hline
\end{tabular}
\end{center}
\end{table}

In a recommender system, each user has voted some objects. Formally,
let $U$ be the set of $m$ users, and $O$ be the set of $n$ objects,
the rating of user $u \in U$ on object $\alpha\in O$ is denoted by
$r_{u\alpha}$. We apply a resource-allocation process (two step of
diffusion) to get similarities between users \cite{Zhou2007,Ou2007}.
Given a user-channel bipartite network (see Fig. 1(b), such a
network is consisted of $m$ users and $5n$ channels), assuming that
a certain amount of resource (e.g., recommendation power) is
associated with each user, we will distribute this resource to other
users via the channels. The process follows two steps. Firstly, each
user distributes his initial resource evenly to all the channels he
connects, and secondly, each channel distributes it's resource
equally to all connected users.

Considering a bipartite graph $G=(U,C,E)$, where $U$ is the set of
users, $C$ is the set of channels, and $E$ is the set of edges
connecting users and channels. After the first step, node $c\in C$
gets the fraction,
\begin{equation}
\label{eq2}
    R_{cv}=\frac{a_{vc}}{k(v)} ,
\end{equation}
of resource from user $v$, where $k(v)$ is the degree of user $v$,
$a_{vc}=1$ if user $v$ is connected to channel $c$, and $a_{vc}=0$
otherwise. Then, at the second step, each channel will distribute
its resource to all the neighboring users. Thus, resource that user
$u$ gets from $v$, defined as the {\it similarity} between $u$ and
$v$, is:
\begin{equation}
\label{eq3} {s_{uv}} = \sum\limits_{c \in C}
\frac{a_{uc}R_{cv}}{k(c)}=\frac{1}{k(v)}\sum\limits_{c \in C}
{\frac{a_{uc}a_{vc}}{k(c)}},
\end{equation}
where $k(c)$ is the degree of channel $c$. Note that, the similarity
matrix $S=(s_{uv})$ is asymmetric, i.e., $s_{uv}\ne s_{vu}$. It is
reasonable because a user who rated a lot of objects often has high
probability to share many common channels with other users and thus
will assign each of them lower weight. Actually, a recent empirical
study showed that this kind of diffusion-based similarity can better
describe the dependence between stations in the Chinese railway
network, comparing with some traditional similarity measures
\cite{Wang2009}. In addition, the whole process obeys the
conservation law, and the similarity matrix is column-normalized, as
$\sum_us_{uv}=1$.

Once we have calculated the user similarities, we can then obtain
the predicted rating on a new object $\alpha\in O$ for a target user
$u\in U$ using the standard collaborative filtering framework, that
is
\begin{equation}\label{eq4}
r'_{u\alpha}=\bar{r}_{u}+\kappa\sum\nolimits_v{s_{uv}}(r_{v\alpha}-\bar{r}_{v}),
\end{equation}
where $\bar{r}_{u}$ denotes the average rating of user $u$,
$\kappa=(\sum_vs_{uv})^{-1}$ serves as the normalization factor, and
$v$ runs over all users having voted the object $\alpha $.

\begin{figure}
  \begin{center}
  \includegraphics[width=4.0in,height=2.0in]{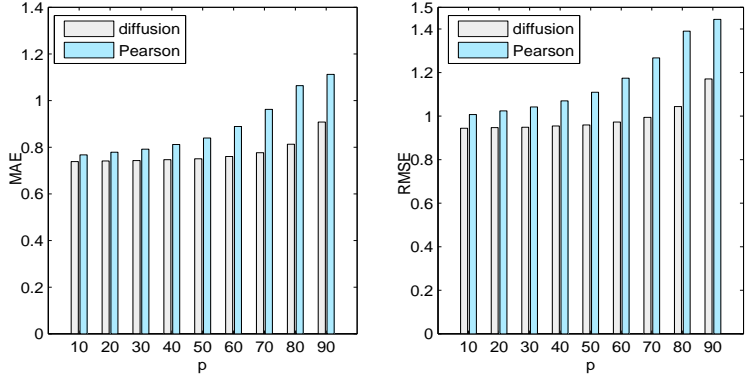}
\caption{Prediction accuracy on \emph{MovieLens} for different
densities of training set. All the number are obtained by averaging
over five runs, each of which has an independently random division
of training set and probe.} \label{fig3}
\end{center}
\end{figure}

\begin{figure}
  \begin{center}
  \includegraphics[width=4.0in,height=2.0in]{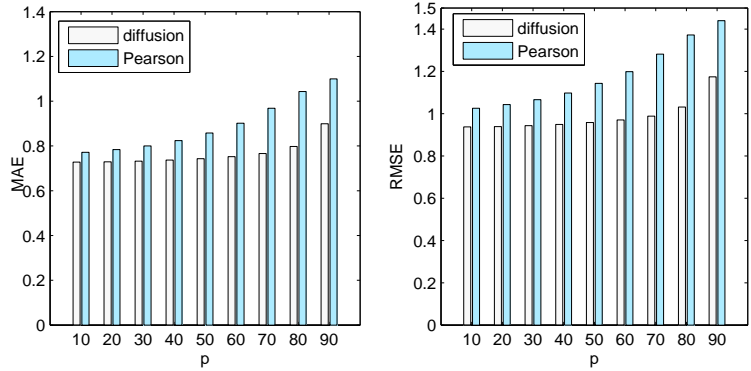}
\caption{Prediction accuracy on \emph{Netflix} for different
densities of training set. All the number are obtained by averaging
over five runs, each of which has an independently random division
of training set and probe.} \label{fig4}
\end{center}
\end{figure}

\section{Numerical results}

To test the algorithmic accuracy, we use two benchmark data sets:
(i)\emph{MovieLens}, which consists of $943$ users, $1682$ objects,
and $10^5$ discrete ratings from 1 to 5. (ii)\emph{Netflix}, which
is a random sample of the original Netflix data set, containing
$3000$ users who have voted at least 45 objects, and $3000$ movies
having been voted at least by 23 users. There are in total $567,456$
ratings. We randomly divide this data set into two parts: one is the
training set, treated as known information, and the other is the
probe, whose information is not allowed to be used for prediction.
we use a parameter, $p\in\{10,20,..,90\}$, to control the data
density, that is, $p\%$ of the ratings are put into the probe set,
and the remains compose the training set.

To evaluate the prediction accuracy, we use two well-known metrics:
\emph{mean absolute error} (MAE) and \emph{root mean square error}
(RMSE). They are respectively defined as:
\begin{equation}\label{errors}
{MAE}=\frac1{\|\mathcal{P}\|}\sum_{(u,\alpha)\in\mathcal{P}} (
r_{u\alpha}- r'_{u\alpha}),
\end{equation}
\begin{equation}
{RMSE}=\sqrt{\frac1{\|\mathcal{P}\|}\sum_{(u,\alpha)\in\mathcal{P}}(r_{u\alpha}-r'_{u\alpha})^2},
\end{equation}
where $\mathcal{P}$ denotes the probe set.

We compare the proposed similarity with a benchmark one, namely the
\emph{Pearson correlation coefficient}, which has been proved highly
competitive to other similarity methods and is widely used in
collaborative filtering algorithms. Under the Pearson's formula, the
similarity, $s_{uv}$, between users $u$ and $v$ is
\begin{equation}
\label{correlation} s_{uv}= \frac{\sum_{\alpha}(r_{u\alpha}-\bar
r_u)(r_{v\alpha}-\bar r_v)} {\sqrt{\sum_{\alpha}(r_{u\alpha}-\bar
r_u)^2} \sqrt{\sum_{\alpha}(r_{v\alpha}-\bar r_v)^2}},
\end{equation}
where $\alpha$ runs over all movies commonly voted by $u$ and $v$.

Table~\ref{tab2} presents the algorithmic accuracies on
\emph{MoiveLens} and \emph{Netflix}. Subject to the prediction
accuracy, one can see that the diffusion-based similarity is notably
better than the classical Pearson correlation coefficient for both
data sets. Figure~\ref{fig3} and Figure~\ref{fig4} report the
comparison between diffusion-based similarity and Pearson
correlation coefficient for different data densities, namely
different $p$. It can be seen that the diffusion-based similarity
outperforms the Pearson correlation coefficient in all cases, and
the difference becomes larger when the data gets sparser, indicating
that this diffusion-based similarity has greater advantage for
sparser systems.

\section{Conclusion and Discussion}
In this paper, by applying a diffusion process, we propose a new
index to quantify the similarity between two users in a user-object
bipartite graph. Under the standard collaborative filtering
framework, we compare the diffusion-based similarity and the
classical Pearson correlation coefficient. The numerical results on
two benchmark data sets, \emph{MovieLens} and \emph{Netflix},
indicated that the diffusion-based similarity can better account for
the proximity of user tastes and provide more accurate predictions.
It is worthwhile to emphasize that the diffusion-based similarity
can give competitively good predictions as the so-called
\emph{transferring similarity} based on Pearson correlation
coefficient \cite{Sun2009}. Since the transferring similarity,
defined as $T=(I-\varepsilon S)^{-1} S$ with $S$ the matrix of
Pearson correlation coefficient and $\varepsilon$ a free parameter,
requires high computational resource and is parameter-dependent, the
diffusion-based similarity, as a local and parameter-free index, is
comparatively more efficient. We think the diffusion-based
similarity, combined with the multi-channel representation, can find
its application especially for the huge-size recommender systems
with discrete ratings.

\section{Acknowledgments}
We acknowledge \emph{GroupLens} Research Group for providing us the
data set \emph{MovieLens}. The \emph{MovieLens} and \emph{Netflix}
data sets can be downloaded from http://www.grouplens.org/ and
http://www.netflixprize.com/, respectively. This work is supported
by the 863 Project (2007AA01Z414), the Chinese Postdoctoral Science
Foundation (20080431273), and the Future and Emerging Technologies
(FET) programme within the Seventh Framework Programme for Research
of the European Commission, under FET-Open grant number 213360
(LIQUIDPUB project). T.Z. acknowledges the National Natural Science
Foundation of China (Grant Nos. 10635040 and 60744003).


\begin{thebibliography}{0}

\bibitem{NJP08} G.-Q. Zhang, G.-Q. Zhang, Q.-F. Yang, S.-Q. Cheng, T.
Zhou, New J. Phys. {\bf 10}, 123027 (2008).

\bibitem{Broder00} A. Broder, R. Kumar, F. Moghoul, P. Raghavan, S. Rajagopalan, R.
Stata, A. Tomkins, J. Wiener, Comput. Netw. {\bf 33}, 309 (2000).

\bibitem{Brin1998} S. Brin, L. Page, Comput. Netw. ISDN Syst. {\bf
30}, 107 (1998).

\bibitem{Kleinberg1999} J. M. Kleinberg, J. ACM {\bf 46}, 604 (1999).

\bibitem{Resnick1997} P. Resnick, H. R. Varian, Commun. ACM {\bf
40}(3), 56 (1997).

\bibitem{Adomavicius05} G. Adomavicius, A. Tuzhilin, IEEE Trans. Knowl. \& Data Eng. {\bf
17}, 734 (2005).

\bibitem{Herlocker04} J. L. Herlocker, J. A. Konstan, K. Terveen, J. T. Riedl, ACM Trans. Inform. Syst. {\bf 22}, 5 (2004).

\bibitem{LiuRev} J.-L. Liu, M. Z. Q. Chen, J. Chen, F. Deng, H.-T.
Zhang, Z.-K. Zhang, T. Zhou, Int. J. Inform. \&. Syst. Sci. {\bf 5},
230 (2009).

\bibitem{Pazzani2007} M. J. Pazzani, D. Billsus, Lect. Notes Comput.
Sci. {\bf 4321}, 325 (2007).

\bibitem{Maslov00} S. Maslov, Y.-C. Zhang, Phys. Rev. Lett. {\bf 87},
248701 (2001).

\bibitem{Goldberg2001} K. Goldberg, T. Roeder, D. Gupta, C. Perkins,
Inf. Retr. {\bf 4}, 133 (2001).

\bibitem{Zhang2007a} Y.-C. Zhang, M. Blattner, Y.-K. Yu, Phys. Rev. Lett. {\bf 99}, 154301 (2007).

\bibitem{Zhang2007b} Y.-C. Zhang, M. Medo, J. Ren, T. Zhou, T. Li, F. Yang, Europhys. Lett. {\bf 80}, 68003 (2007).

\bibitem{Zhou2007} T. Zhou, J. Ren, M. Medo, Y.-C. Zhang, Phys. Rev. E {\bf 76}, 046115 (2007).

\bibitem{Zhou2008} T. Zhou, L.-L. Jiang, R.-Q. Su, Y.-C. Zhang, Europhys. Lett. {\bf 81}, 58004 (2008).

\bibitem{Hofmann2004} T. Hofmann, ACM Trans. Inform. Syst. {\bf 22}, 89 (2004).

\bibitem{Blei2003} D. M. Blei, A. Y. Ng, M. I. Jordan, J. Mech.
Learn. Res. {\bf 3}, 993 (2003).

\bibitem{Ren2008} J. Ren, T. Zhou, Y.-C. Zhang, Europhys. Lett. {\bf 82}, 58007
(2008).

\bibitem{Schafer2007} J. B. Schafer, D. Frankowski, J. L. Herlocker,
S. Sen, Lect. Notes Comput. Sci. {\bf 4321}, 291 (2007).

\bibitem{Fouss2007} F. Fouss, A. Pirotte, J.-M. Renders, M. Saerens,
IEEE Trans. Knowl. \& Data. Eng. {\bf 19}, 355 (2007).

\bibitem{Linden2003} G. Linden, B. Smith, J. York, IEEE Internet
Comput. {\bf 7}, 76 (2003).

\bibitem{Lambiotte} R. Lambiotte, M. Ausloos, Phys. Rev. E {\bf 72},
066107 (2005).

\bibitem{Salton1983} G. Salton, M. J. McGill, \emph{Introduction to Modern
Information Retrieval} (MuGraw-Hill, Auckland, 1983).

\bibitem{Gobel1974} F. Gobel, A. Jagers, Stochastic Processes and
Their Applications {\bf 2}, 311 (1974).

\bibitem{Huang2004} Z. Huang, H. Chen, D. Zeng, ACM Trans. Inform. Syst. {\bf 22}, 116 (2004).

\bibitem{Liu2009} R.-R. Liu, C.-X. Jia, T. Zhou, D. Sun, B.-H. Wang,
Physica A {\bf 388}, 462 (2009).

\bibitem{LiuJG2009} J.-G. Liu, B.-H. Wang, Q. Guo, Int. J. Mod.
Phys. C {\bf 20}, 285 (2009).

\bibitem{Sun2009} D. Sun, T. Zhou, R.-R. Liu, C.-X. Jia, J.-G. Liu,
B.-H. Wang, arXiv: 0807.4495.

\bibitem{Katz1953} L. Katz, Psychmetrika {\bf 18}, 39 (1953).

\bibitem{Chebotarev1997} P. Chebotarev, E. Shamis, Automation and
Remote Control {\bf 58}, 1505 (1997).

\bibitem{Ou2007} Q. Ou, Y.-D. Jin, T. Zhou, B.-H. Wang, B.-Q. Yin,
Phys. Rev. E {\bf 75}, 021102 (2007).

\bibitem{Wang2009} Y.-L. Wang, T. Zhou, J.-J. Shi, J. Wang, D.-R.
He, Physica A {\bf 388}, 2949 (2009).

\end{thebibliography}
\end{document}